\documentclass[12pt]{article}

\pdfoutput=1

\usepackage{graphics}
\usepackage{epsfig}
\usepackage{amsfonts}
\usepackage{amssymb}
\usepackage{amsmath}
\usepackage{dsfont}
\usepackage{pifont}
\usepackage{bbm}
\usepackage{multirow}
\usepackage{latexsym}
\usepackage{verbatim}
\usepackage{mcite}
\usepackage{cite}
\usepackage{units}
\usepackage[all]{xy}
\usepackage{cancel}
\usepackage{slashed}

\def\a{{\alpha}}
\def\b{{\beta}}

\def\g{{\gamma}}

\def\d{{\delta}}

\def\bfone{\relax{\rm 1\kern-.35em 1}}

\newcommand{\cN}{{\cal N}}

\newcommand{\be}{\begin{equation}}
\newcommand{\ee}{\end{equation}}
\newcommand{\ben}{\begin{displaymath}}
\newcommand{\een}{\end{displaymath}}
\newcommand{\bea}{\begin{eqnarray}}
\newcommand{\eea}{\end{eqnarray}}
\newcommand{\nn}{\nonumber}

\newcommand{\bean}{\begin{eqnarray*}}
\newcommand{\eean}{\end{eqnarray*}}

\DeclareMathAlphabet{\mathpzc}{OT1}{pzc}{m}{it}

\begin{document}
\pagestyle{plain}


\makeatletter \@addtoreset{equation}{section} \makeatother
\renewcommand{\thesection}{\arabic{section}}
\renewcommand{\theequation}{\thesection.\arabic{equation}}
\renewcommand{\thefootnote}{\arabic{footnote}}


\setcounter{page}{1} \setcounter{footnote}{0}


\begin{titlepage}

\begin{flushright}
UUITP-27/12\\
\end{flushright}

\bigskip

\begin{center}

\vskip 0cm

{\LARGE \bf On the distribution of stable de Sitter vacua} \\[6mm]

\vskip 0.5cm

{\bf Ulf Danielsson \,and\, Giuseppe Dibitetto}\\

\vskip 25pt

{\em Institutionen f\"or fysik och astronomi, \\ 
University of Uppsala, \\ 
Box 803, SE-751 08 Uppsala, Sweden \\
{\small {\tt \{ulf.danielsson, giuseppe.dibitetto\}@physics.uu.se}}} \\

\vskip 0.8cm

\end{center}

\vskip 1cm

\begin{center}

{\bf ABSTRACT}\\[3ex]

\begin{minipage}{13cm}
\small

The possible existence of (meta-) stable de Sitter vacua in string theory is of fundamental importance. So far, 
there are no fully stable solutions where all effects are under perturbative control. 
In this paper we investigate the presence of stable de Sitter vacua in type II string theory with non-geometric fluxes.
We introduce a systematic method for solving the
equations of motion at the origin of moduli space, by expressing the fluxes in terms of the supersymmetry breaking 
parameters. As a particular example, we revisit the geometric type IIA compactifications, and argue 
that non-geometric fluxes are necessary to have (isotropically) stable de Sitter solutions.
We also analyse a class of type II compactifications with non-geometric fluxes, and study the distribution of (isotropically) 
stable de Sitter points in 
the parameter space. We do this through a random scan as well as through a complementary analysis of
two-dimensional slices of the parameter
space. We find that the (isotropically) stable de Sitter vacua are surprisingly rare, and organise themselves into 
thin sheets at small values of the cosmological constant.

\end{minipage}

\end{center}

\vfill

\end{titlepage}


\tableofcontents

\section{Introduction}
\label{sec:introduction}

Since the turn of the millennium, there have been many attempts to embed dark energy 
and inflation in string theory using flux compactifications. 
Parallelly, a whole class of string theory constructions involving flux backgrounds compatible with minimal supersymmetry 
has been considered in the literature. In particular, the mechanism of inducing effective 
superpotentials from fluxes \cite{Giddings:2001yu} has often been considered in the 
context of the supergravity descriptions normally referred to as $STU$-models 
\cite{Kachru:2002he, Derendinger:2004jn, DeWolfe:2004ns, Camara:2005dc, Villadoro:2005cu, Derendinger:2005ph, DeWolfe:2005uu, Aldazabal:2006up, Aldazabal:2007sn}. It has, 
however, turned out to be surprisingly difficult to find the required stable or quasi-stable 
de Sitter (dS) solutions through compactifications of string theory. Let us start with a brief review 
of the main results of these searches.
    
Based on the results of~\cite{Flauger:2008ad}, in ref.~\cite{Caviezel:2008tf} all the type IIA coset 
compactifications with $\textrm{SU}(3)$ structure were analysed, and a no-go result was proven in all the cases but one. 
The only model evading these no-go theorems turned out to be an $\textrm{SU}(2) \times \textrm{SU}(2)$ compactification. 
This model was indeed numerically shown to contain a dS solution, which was found to be unstable due to the 
presence of a tachyon already within the isotropic subsector. This example of a type IIA unstable dS extremum was 
later studied in ref.~\cite{Danielsson:2010bc} (see also \cite{Danielsson:2009ff}),  where it was shown to belong to a whole continuous line of tachyonic solutions, for which the ten-dimensional uplift to massive IIA supergravity 
was constructed in terms of very simple torsion classes. 
This was accommodated in a general framework in \cite{Danielsson:2011au}. For further analysis of this model, see ref.~\cite{Danielsson:2012et}.
In parallel, in ref.~\cite{deCarlos:2009fq}, many type II setups in the $\mathbb{Z}_{2} \,\times\, \mathbb{Z}_{2}$ orbifold with generalised fluxes were analysed searching 
for dS solutions. In a setup corresponding to type IIA with O$6$/D$6$ and only geometric fluxes, a dS region was found in the 
neighbourhood of a Minkowski critical point, but again with no room for stability. Later in ref.~\cite{Dibitetto:2010rg}, 
the original dS solution of ref.~\cite{Caviezel:2008tf} was localised at a particular point within the 
aforementioned dS region.
    
Finally in ref.~\cite{Dibitetto:2011gm}, the complete set of solutions in the framework of isotropic $\cN=1$ 
geometric type IIA compactifications on $T^{6}/(\mathbb{Z}_{2}\,\times\,\mathbb{Z}_{2})$ was presented. It turns out to consist of a discrete set of lines, of which only one 
crosses the Minkowski point \cite{deCarlos:2009qm} and enters the dS region, with tachyons at every point. This can be regarded as an indication that one 
needs to consider more general setups including non-geometric fluxes \cite{Shelton:2005cf} in order to find stable dS vacua \cite{deCarlos:2009fq}.
    
The main goal of this paper is to show how non-geometric fluxes indeed provide enough freedom in parameter space for tuning the masses of all the moduli to be 
positive. Moreover, in situations where this is 
the case, we would like to locate the region of stable dS in parameter space and qualitatively estimate which portion of the total volume this corresponds to.
    
The paper is organised as follows. In section~\ref{sec:review} we introduce the $\mathbb{Z}_{2} \,\times\, \mathbb{Z}_{2}$ 
compactifications with orientifold and their superpotential formulations. Subsequently, we 
present a new technique for solving the equations of motion in the origin of moduli space. 
In such a framework it becomes straightforward to predict the dimensionality of the parameter space of solutions. 
Finally, we make use of this machinery to review type IIA compactifications with geometric fluxes regarded as an over-constrained particular sub-case. 
In section~\ref{sec:non_geom_IIB} a IIB setup with non-geometric fluxes is introduced in which the parameter space has the correct degrees of freedom 
to gain control of the masses for all the isotropic moduli. A statistical scan of the parameter space is performed, and the fraction of stable solutions 
is plotted as a function of the normalised cosmological constant. Furthermore, we show some two-dimensional 
slices of the parameter space, and locate the thin region of stable dS critical points. 
Finally, we present our conclusions. An appendix is provided where we give some details about the analytical solutions in type IIB with $F$, $H$ and $Q$ fluxes.

\section{$\cN=1$ compactifications with generalised fluxes}
\label{sec:review}

\subsection{The $\mathbb{Z}_{2}\,\times\,\mathbb{Z}_{2}$ orbifold}

A wide class of string compactifications down to four dimensions compatible with minimal supersymmetry can be effectively 
described by particular supergravity theories which are known as $STU$-models. These theories arise, \emph{e.g.}, from $T^{6}/(\mathbb{Z}_{2}\,\times\,\mathbb{Z}_{2})$ orbifold compactifications of type IIB with O3/O7-planes (and duals thereof). This class of string compactifications has the very appealing feature of collecting together low-energy effective descriptions which are related to each other by dualities without the intervention of mirror symmetry. In the $\mathbb{Z}_{2}\,\times\,\mathbb{Z}_{2}$ orbifold any dual effective theory is still formally described by the same Lagrangian where only fields and couplings have been transformed.

This class of effective descriptions enjoys $\cN=1$ supersymmetry and $\textrm{SL}(2)^{7}$ global bosonic symmetry due to the coupling of the gravity multiplet with seven extra chiral multiplets. The scalar sector contains seven complex fields spanning the coset space $\left(\textrm{SL}(2)/\textrm{SO}(2)\right)^{7}$ which we denote by $\Phi^{\a}\,\equiv\,\left(S,T_{i},U_{i}\right)$ with $i=1,2,3$. In the type IIB language, the $S$ modulus contains the ten-dimensional dilaton, whereas the $T$ and $U$ moduli are interpreted as K\"ahler and 
complex structure moduli respectively\footnote{Please note that, in a type IIA language, the ten-dimensional dilaton would correspond to a combination of $S$ and $T$, whereas the K\"ahler and 
complex structure moduli would be interchanged.}.
The kinetic Lagrangian can be derived from the following K\"ahler potential
\be
\label{Kaehler_STU}
K\,=\,-\log\left(-i\,(S-\overline{S})\right)\,-\,\sum_{i=1}^{3}{\log\left(-i\,(T_{i}-\overline{T}_{i})\right)}\,-\,\sum_{i=1}^{3}{\log\left(-i\,(U_{i}-\overline{U}_{i})\right)}\ .
\ee
This yields
\be
\mathcal{L}_{\textrm{kin}} = \frac{\partial
S\partial \overline{S}}{\left(-i(S-\overline{S})\right)^2} + \, \sum_{i=1}^{3}\left(\frac{\partial
T_{i}\partial \overline{T}̣_{i}}{\left(-i(T_{i}-\overline{T}_{i})\right)^2} + \frac{\partial
U_{i}\partial \overline{U}_{i}}{\left(-i(U_{i}-\overline{U}_{i})\right)^2}\right) \ .
\ee

The presence of fluxes induces a scalar potential $V$ for the moduli fields which is given in terms of the above K\"ahler potential and a holomorphic superpotential $W$ by
\be
\label{V_N=1}
V\,=\,e^{K}\left(-3\,|W|^{2}\,+\,K^{\a\bar{\b}}\,D_{\a}W\,D_{\bar{\b}}\overline{W}\right)\ ,
\ee
where $K^{\a\bar{\b}}$ is the inverse K\"ahler metric and $D$ denotes the K\"ahler-covariant derivative.
 
As a first example of flux-induced deformation, let us consider the presence of gauge fluxes in the duality frame of type IIB with O$3$ and O$7$-planes which gives rise to the following superpotential
\be
\label{W_GKP}
W\,=\underbrace{\,P_{F}(U_{i})\,}_{F \textrm{ flux}}\,+\,\underbrace{\,P_{H}(U_{i})\,S\,}_{H \textrm{ flux}}\ ,
\ee
where $P_{F}$ and $P_{H}$ are cubic polynomials in the complex structure moduli given by
\be
\begin{array}{cclc}
P_{F}(U_{i}) & = & a_{0}\,-\,\sum\limits_{i}{a_{1}^{(i)}\,U_{i}}\,+\,\sum\limits_{i}{a_{2}^{(i)}\,\dfrac{U_{1}\,U_{2}\,U_{3}}{U_{i}}}\,-\,a_{3}\,U_{1}\,U_{2}\,U_{3} & , \\[3mm]
P_{H}(U_{i}) & = & -b_{0}\,+\,\sum\limits_{i}{b_{1}^{(i)}\,U_{i}}\,-\,\sum\limits_{i}{b_{2}^{(i)}\,\dfrac{U_{1}\,U_{2}\,U_{3}}{U_{i}}}\,+\,b_{3}\,U_{1}\,U_{2}\,U_{3} & . 
\end{array}
\ee
The $\cN=1$ supergravity defined by the above superpotential has a \emph{no-scale} feature due to the absence of the moduli $T_{i}$. This implies that their real parts appear as completely flat directions in the scalar potential. The line of including some non-perturbative effects such as gaugino condensation \cite{Font:1990nt} has been considered as a possible mechanism to further stabilise the K\"ahler moduli (see \emph{e.g.} refs~\cite{Witten:1996bn, Achucarro:2006zf}).

Duality covariance arguments would suggest to complete the superpotential given in \eqref{W_GKP} with more effective couplings which are obtained by performing $\textrm{SL}(2)^{7}$ transformations. This gives rise to a generalised form of $W$ containing all the possible $2^{7}=128$ couplings from the constant one up to the $ST_{1}T_{2}T_{3}U_{1}U_{2}U_{3}$ term. Such a very general superpotential, though, will be non-geometric in any duality frame. 

\subsection*{The isotropic truncation}

For the sake of simplicity, in the main part of this paper we will restrict ourselves to the isotropic limit of the theories previously presented. This reduces the number of complex moduli down to three via
\be
\begin{array}{lclc}
T_{1}\,=\,T_{2}\,=\,T_{3}\,\equiv\,T & \textrm{ and } & U_{1}\,=\,U_{2}\,=\,U_{3}\,\equiv\,U & ,
\end{array}
\ee
which span the scalar coset $\left(\dfrac{\textrm{SL}(2)}{\textrm{SO}(2)}\right)^{3}$. As for the fluxes, taking the isotropic limit implies the following identifications
\be
a_{1}^{(1)}\,=\,a_{1}^{(2)}\,=\,a_{1}^{(3)}\,\equiv\,a_{1}
\ee
and similarly for $a_{2}^{(i)}$, $b_{1}^{(i)}$ and $b_{2}^{(i)}$.

In the isotropic limit, the complete duality covariant superpotential contains $32$ couplings going from the constant term up to the $ST^{3}U^{3}$ term. 
For completeness, we give here the full form of the flux-induced $W$ and its stringy interpretation in the different duality frames. 

\begin{figure}[h!]
\begin{center}
\scalebox{0.9}[0.9]{
\begin{tabular}{ccccc}
\includegraphics[scale=0.5,keepaspectratio=true]{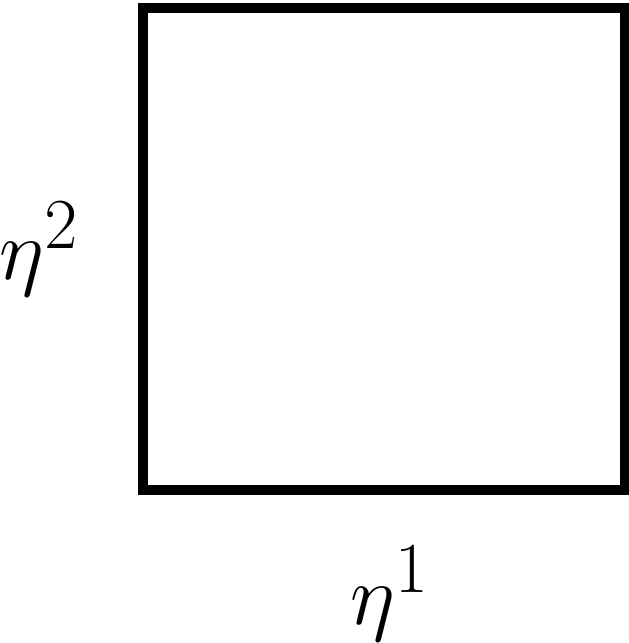} &    &  \includegraphics[scale=0.5,keepaspectratio=true]{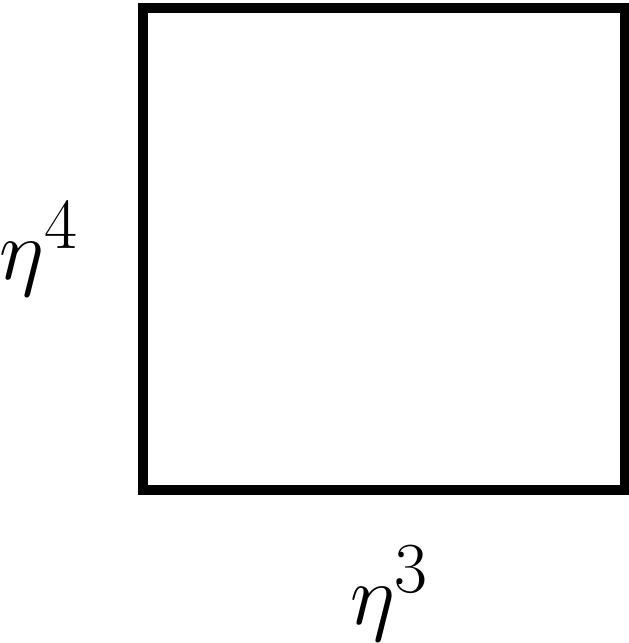}  &    & \includegraphics[scale=0.5,keepaspectratio=true]{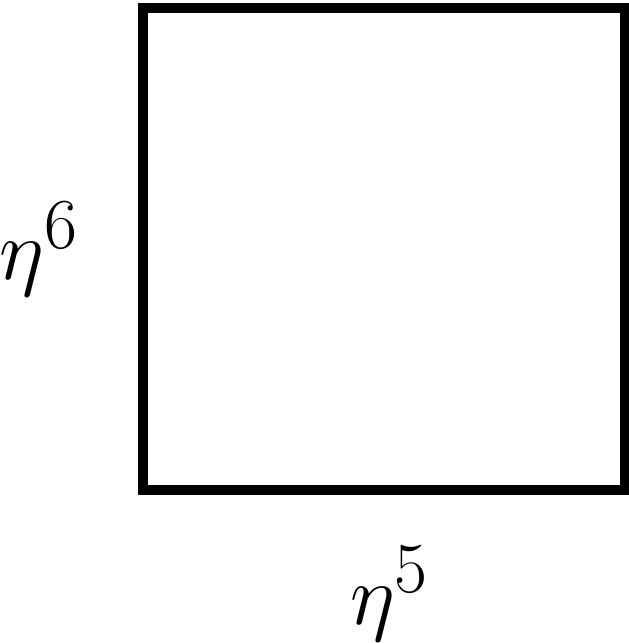} \\[-24mm]
     & \,\,\Large{$\times$} &        & \,\,\Large{$\times$} &
\end{tabular}
}
\end{center}
\vspace{9.5mm}
\caption{{\it $T^{6} =  T^{2}_{1} \times T_{2}^{2} \times T_{3}^{2}$ torus factorisation and the coordinate basis. We will use indices $a,b,c$ for horizontal $\,``-"$ directions $1,\,3,\,5$ and indices $i,j,k$ for vertical $\,``\,|\,"$ directions $2,\,4,\,6$ in the $2$-tori $\,T_{i}\,$ with $\,i=1,2,3$.}}
\label{fig:Torus_Factor1}
\end{figure}

The orbifold involution forces the six-dimensional internal space to be factorised into three $2$-tori as shown in figure~\ref{fig:Torus_Factor1}.
Now we will summarise the correspondence between generalised isotropic fluxes and superpotential couplings appearing in the $\cN=1$ effective description. 
The generalised flux-induced superpotential can be written as 
\be
\label{W_all_fluxes}
W = (P_{F} + P_{H} \, S ) + 3 \, T \, (P_{Q} + P_{P} \, S ) + 3 \, T^2 \, (P_{Q'} + P_{P'} \, S ) + T^3 \, (P_{F'} + P_{H'} \, S ) \ ,
\ee
where\footnote{\label{c1_tilde}Please note that, in principle, the truncation to the isotropic sector gives rise to $32+8=40$ fluxes, where all the fluxes transforming in the mixed symmetry representations of $\textrm{GL}(6)$ (\emph{i.e.} $Q$, $P$ and their primed counterparts in type IIB) have in fact two fluxes $(c_{1},\tilde{c}_{1})$ etc. giving rise to one single coupling $(2c_{1}-\tilde{c}_{1})$ etc., so without loss of generality, we set throughout the text $\tilde{c}_{1}=c_{1}$ etc..}
\be
\begin{array}{lcll}
\label{Poly_unprim}
P_{F} = a_0 - 3 \, a_1 \, U + 3 \, a_2 \, U^2 - a_3 \, U^3 & \hspace{5mm},\hspace{5mm} & P_{H} = b_0 - 3 \, b_1 \, U + 3 \, b_2 \, U^2 - b_3 \, U^3 & ,  \\[2mm]
P_{Q} = c_0 + c_{1} \, U - c_{2} \, U^2 - c_3 \, U^3 & \hspace{5mm},\hspace{5mm} & P_{P} = d_0 + d_{1} \, U - d_{2} \, U^2 - d_3 \, U^3 & ,
\end{array}
\ee
and
\be
\begin{array}{lcll}
\label{Poly_prim}
P_{F'} = a_3' + 3 \, a_2' \, U + 3 \, a_1' \, U^2 + a_0' \, U^3 & \hspace{3mm},\hspace{3mm} &P_{H'} = b_3' + 3 \, b_2' \, U + 3 \, b_1' \, U^2 + b_0' \, U^3 & ,  \\[2mm]
P_{Q'} = -c_3' +  c'_{2} \, U + c'_{1} \, U^2 - c_0' \, U^3 & \hspace{3mm},\hspace{3mm} & P_{P'} = -d_3' + d'_{2} \, U + d'_{1} \, U^2 - d_0' \, U^3 & .
\end{array}
\ee
The first half of the couplings represent fluxes which admit a locally geometric interpretation in type IIB (unprimed fluxes), whereas the remaining ones (primed fluxes) represent additional generalised fluxes which do not even admit a locally geometric description. These were first formally introduced in ref.~\cite{Aldazabal:2006up} as dual counterparts of the unprimed fluxes. 

\begin{table}[h!]
\renewcommand{\arraystretch}{1.25}
\begin{center}
\scalebox{0.92}[0.92]{
\begin{tabular}{ | c || c | c | c || c |}
\hline
couplings & Type IIB & Type IIA & fluxes & non-iso \emph{dof}'s\\
\hline
\hline
$1 $&  $ {F}_{ ijk} $& $F_{aibjck}$ & $  a_0 $ & $1$\\
\hline
$U $&  ${F}_{ ij c} $& $F_{aibj}$ & $   a_1 $ & $3$\\
\hline
$U^2 $& ${F}_{i b c} $& $F_{ai}$ & $  a_2 $ & $3$\\
\hline
$U^3 $& ${F}_{a b c} $& $F_{0}$ & $  a_3 $ & $1$\\
\hline
\hline
$S $& $ {H}_{ijk} $& $ {H}_{ijk} $  & $  -b_0$ & $1$\\
\hline
$S \, U $& ${H}_{ij c} $& ${{\omega}_{ij}}^{c}$ & $  -b_1 $ & $3$\\
\hline
$S \, U^2 $&  ${H}_{ i b c}$ & $ {{Q}_{ i }}^{ b c}$  & $  -b_2 $ & $3$\\
\hline
$S \, U^3 $& $ {H}_{a b c} $& $ {R}^{a b c} $ & $  -b_3 $ & $1$\\
\hline
\hline
$T $& $  {Q_k}^{a b} $&$ H_{a b k} $ & $  c_0 $ & $3$\\
\hline
$T \, U $& $ {Q_k} ^{a j} = {Q_k}^{i b} \,\,\,,\,\,\, {Q_a}^{b c} $& $ {\omega_{k a}}^{j} = {\omega_{b k}}^{i} \,\,\,,\,\,\, {\omega_{b c}}^a $  & $c_1 \,\,\,,\,\,\, \tilde {c}_1 $ & $9$\\
\hline
$T \, U^2 $& $ {Q_c}^{ib} = {Q_c}^{a j} \,\,\,,\,\,\, {Q_k}^{ij} $& $ {Q_b}^{ci} = {Q_a}^{j c} \,\,\,,\,\,\, {Q_k}^{ij} $ & $c_2 \,\,\,,\,\,\,\tilde{c}_2 $ & $9$\\
\hline
$T \, U^3 $& $  {Q_{c}}^{ij} $& $  R^{ijc} $ & $c_3 $ & $3$\\
\hline
\hline
$S \, T $& $ {P_k}^{a b}$ & & $  -d_0 $ & $3$\\
\hline
$S \, T \, U $& $ {P_k}^{a j} = {P_k}^{i b} \,\,\,,\,\,\, {P_a}^{b c} $&  & $-d_1 \,\,\,,\,\,\, -\tilde{d}_1 $ & $9$\\
\hline
$S \, T \, U^2 $& $ {P_c}^{ib}= {P_c}^{a j} \,\,\,,\,\,\, {P_k}^{ij} $&  & $-d_2 \,\,\,,\,\,\,-\tilde{d}_2 $ & $9$\\
\hline
$S \, T \, U^3 $& $  {P_{c}}^{ij} $&  & $-d_3 $ & $3$\\
\hline
\end{tabular}
}
\end{center}
\caption{{\it Mapping between unprimed fluxes and couplings in the superpotential both in type IIB with O3 and O7 and in type IIA with O6. The six internal directions depicted in figure~\protect\ref{fig:Torus_Factor1} are split into $\,``-"$ labelled by $i=1,3,5$ and $\,``\,|\,"$ labelled by $a=2,4,6$. Note that the empty boxes in type IIA are related to the presence of dual fluxes analogous to the 'primed' notation in type IIB.
The last column refers to the number of independent superpotential couplings the corresponding fluxes give rise to once the non-isotropic moduli are opened up.}}
\label{table:unprimed_fluxes}
\end{table}

\begin{table}[h!]
\renewcommand{\arraystretch}{1.25}
\begin{center}
\scalebox{0.92}[0.92]{
\begin{tabular}{ | c || c | c | c || c |}
\hline
couplings &  Type IIB &  Type IIA & fluxes & non-iso \emph{dof}'s\\
\hline
\hline
$T^3 \, U^3 $& $ {F'}^{ijk} $&  & $  a_0' $ & $1$\\
\hline
$T^3 \, U^2 $& ${F'}^{ ij c} $& &$   a_1' $ & $3$\\
\hline
$T^3 \, U $& ${F'}^{i b c} $& &$  a_2' $ & $3$\\
\hline
$ T^3 $& ${F'}^{a b c} $& &$  a_3' $ & $1$\\
\hline
\hline
$S \, T^3 \, U^3 $& $ {H'}^{ ijk} $& &$  -b_0'$ & $1$\\
\hline
$S \, T^3 \, U^2 $& $ {H'}^{i jc} $& &$ - b_1' $ & $3$\\
\hline
$S \, T^3 \, U $& $ {H'}^{ i b c} $& & $  -b_2' $ & $3$\\
\hline
$S  \, T^3 $& $ {H'}^{a b c} $& &$  -b_3' $ & $1$\\
\hline
\hline
$T^2 \, U^3 $& $  {{Q'}_{a b}}^k $& &$  c_0' $ & $3$ \\
\hline
$T^2 \, U^2 $& $ {{Q'}_{a j}}^k = {{Q'}_{i b}}^k \,\,\,,\,\,\, {{Q'}_{b c}}^a $& &$c_1' \,\,\,,\,\,\, \tilde{c}_1' $ & $9$\\
\hline
$T^2 \, U $& $ {{Q'}_{ib}}^c = {{Q'}_{a j}}^c \,\,\,,\,\,\, {{Q'}_{ij}}^k $& &$c_2' \,\,\,,\,\,\,\tilde{c}_2' $ & $9$\\
\hline
$T^2 $& $ {{Q'}_{ij}}^{c} $& &$c_3' $ & $3$\\
\hline
\hline
$S \, T^2 \, U^3$& $  {{P'}_{a b}}^k $& &$  -d_0' $ & $3$\\
\hline
$S \, T^2 \, U^2 $& $ {{P'}_{a j}}^k = {{P'}_{i b}}^k \,\,\,,\,\,\, {{P'}_{b c}}^a $& &$-d_1' \,\,\,,\,\,\, -\tilde{d}_1' $ & $9$\\
\hline
$S \, T^2 \, U $& $ {{P'}_{ib}}^c = {{P'}_{a j}}^c \,\,\,,\,\,\, {{P'}_{ij}}^k $& &$-d_2' \,\,\,,\,\,\,-\tilde{d}_2' $ & $9$\\
\hline
$S \, T^2  $& $  {{P'}_{ij}}^{c} $& & $-d_3' $ & $3$\\
\hline
\end{tabular}
}
\end{center}
\caption{{\it Mapping between primed fluxes and couplings in the superpotential. The conventions are the as in table~\protect\ref{table:unprimed_fluxes} and again, just as there, the empty column should be filled in with extra dual fluxes.}}
\label{table:primed_fluxes}
\end{table}

For our present purposes we focus on superpotentials induced by generalised fluxes in type IIB with O$3$ and O$7$-planes which describe a locally geometric string background. Such a form of the superpotential turns out to be given by
\be
\label{W_IIB_non_geom}
W^{(\textrm{loc.~geom.})}\,=\,(P_{F}\,+\,P_{H} \, S ) \,+\, 3 \, T \, (P_{Q} \,+\, P_{P} \, S )\ ,
\ee
where each polynomial $P_{F}$, $P_{H}$, $P_{Q}$ and $P_{P}$ has degree three with respect to $U$. Their explicit form and their interpretation in terms of the IIB fluxes are respectively given in 
\eqref{Poly_unprim} and table~\ref{table:unprimed_fluxes}.

In this paper we will be interested in constraining the stability of solutions of these isotropic $STU$-models originating from (generalised) string compactifications. The first relevant constraint comes from the analysis
of the supersymmetry breaking sector and its universal features which allow one to constrain metastable de Sitter vacua \cite{Covi:2008cn} and inflation \cite{Borghese:2012yu, Achucarro:2012hg}. 
The averaged projection of the scalar mass matrix along the supersymmetry breaking directions results in the following sGoldstino bound for the $\eta$ parameter
\be
\label{sGoldstino_bound}
\begin{array}{lclclc}
\eta & \equiv & \textrm{Min}\left\{\textrm{Eigenvalues}\left({\left(m^{2}\right)^{I}}_{J}\right)\right\} & \le & \dfrac{2}{3\g} \,-\, \dfrac{1+\g}{\g}\,\tilde{\mathcal{R}} & ,
\end{array}
\ee
where $\g\,\equiv\,\frac{V}{3\,e^{K}\,|W|^{2}}$ and $\tilde{\mathcal{R}}$ denotes the sectional curvature, \emph{i.e.} the projection of the Riemann tensor along the sGoldstino.

In the present supergravity setup we can explicitely evaluate the range of variation of the sectional curvature term in \eqref{sGoldstino_bound}. $\tilde{\mathcal{R}}$ can be written as $2/n_{\textrm{eff}}$, where 
\be
n_{\textrm{eff}} \,=\, \frac{\left(G_{\a}G^{\a}\right)^{2}}{\left(G_{S}G^{S}\right)^{2}+\frac{1}{3}\left(G_{T}G^{T}\right)^{2}+\frac{1}{3}\left(G_{U}G^{U}\right)^{2}} \ ,
\ee
where $G$ denotes the K\"ahler-invariant combination \cite{Cremmer:1978hn} $K \,+\, \log |W|^{2}$ and its first derivative $G_{\a}$ is normalised such that $G_{\a}G^{\a}=3(1+\g)$. $n_{\textrm{eff}}$ gives a notion of how many fields 
effectively participate in the supersymmetry breaking phenomenon. Such a quantity turns out to be a real number 
between $1$ and $7$, these two extremal cases corresponding, respectively, to the limits in 
which either $G_{\a}$ only points in the $S$ direction or $G_{\a} \sim (1,3,3)$\footnote{The $T$ and $U$ components 
are equal to $3$ instead of $1$ because this model originates from the $\textrm{SL}(2)^{7}$ 
theory in the isotropic limit, where the triplets of complex moduli $T_{i}$ and $U_{i}$ collapse into $T$ and $U$ respectively.}. 

The sGoldstino bound in \eqref{sGoldstino_bound} implies in this case that, whenever $V>0$, 
\be
\label{new_sG_bound}
n_{\textrm{eff}} \,\geq\, 3 \, (1+\g) \ ,
\ee
which roughly tells us that stable dS favours homogeneous supersymmetry breaking rather than the other extremal situation in which it takes places mostly in one complex direction.

\subsection{A new systematic prescription}
\label{subsec:counting}

Now we will present a general method for symplifying and subsequently solving the field equations in the origin of moduli space. If the scalar manifold is homogenous and the superpotential 
couplings constitute a closed set under non-compact duality transformations\footnote{Please note that the homogeneity condition is satisfied in supergravity theories arising from $\mathbb{Z}_{2} \,\times\, \mathbb{Z}_{2}$ orbifold compactifications, both in the isotropic and in the non-isotropic case.
As for the second assumption, it turns out to be true in the two examples which are relevant here, \emph{i.e.} type IIA with geometric fluxes and type IIB with generalised fluxes.}, 
restricting the search for solutions to the origin of moduli space does not imply any loss of generality. 
This argument was first employed in ref.~\cite{Dibitetto:2011gm} in the context of $\cN=4$ and $\cN=1$ supergravity. 
The advantage of this is that the equations of motion for the scalar fields turn into a system of quadratic conditions for the fluxes.

Our prescription relies on a linear change of variables which introduces a number $N$ of supersymmetry breaking parameters, $N$ being the number of real fields in the model. 
The ``broken'' supersymmetry equations in the origin $\Phi_{0}$ for the complex moduli $\left\{\Phi^{\a}\right\}_{\a=1,\,\dots,\,N/2}$ read
\be
\label{SUSY_break}
D_{\a}W|_{\Phi_{0}}=A_{\a} \,+\, iB_{\a} \ , 
\ee
where $A_{\a}$ and $B_{\a}$ denote $N$ real constants and the lhs of \eqref{SUSY_break} is a linear expression in the $M$ fluxes. This implies that, whenever $M>N$, the above equations can always be solved in terms of $N$ fluxes as a linear function of the supersymmetry breaking parameters.

By plugging the expression of the solution of \eqref{SUSY_break} into the field equations in the origin, one obtains a new system of equations in which the new unknowns are the $N$ supersymmetry breaking parameters and the remaining $(M-N)$ fluxes.
The peculiar feature of this new system lies in the fact that the extra fluxes will now appear only linearly in the field equations, the only unknows which still appear quadratically being $A_{\a}$ and $B_{\a}$.
If one arranges the supersymmetry breaking parameters into a real vector $X_{I}$ with $I=1,\,\dots,\,N$ and denotes the fluxes by $\left\{F^{A}\right\}_{A=1,\,\dots,\,M}$, the field equations will take the form 
\be
\label{EOM_trick}
h^{IJ}\,X_{I}\,X_{J} \,+\, {g^{I}}_{A}\,X_{I}\,F^{A} \,=\, 0 \ , 
\ee
where $h$ and $g$ denote some real constants.
The reason why this always has to be the case is that the supersymmetry equations should imply the field equations. Hence, whenever the rhs of \eqref{SUSY_break} vanishes there cannot be any quadratic conditions left to be imposed on the remaining fluxes.

This makes it particularly simple to solve the field equations in the origin in the case in which the extra fluxes remaining after the introduction of the supersymmetry breaking parameters are at least $N$ (this requires $M \overset{!}{\geq} 2N$).
This sequence can be sketchily represented as
\be
\hspace{-5mm}
\begin{array}{lclclc}
\left(\begin{array}{c} F^{1} \\ \vdots \\ F^{M}\end{array}\right) & \overset{\eqref{SUSY_break}}{\rightarrow} & \left(\begin{array}{c} X_{1} \\ \vdots \\ X_{N} \\ F^{N+1} \\ \vdots \\ F^{M}\end{array}\right) & \overset{\eqref{EOM_trick}}{\rightarrow} 
& \left(\begin{array}{c} F^{1} \\ \vdots \\ F^{2N}\end{array}\right) = \left(\begin{array}{c} F^{1} \\ \vdots \\ F^{2N}\end{array}\right)(X_{1},\,\dots,\,X_{N},\,\,F^{2N+1},\,\dots,\,F^{M}) & .
\end{array}
\nn
\ee

\begin{figure}[h!]
\begin{center}
\scalebox{1}[0.9]{
\includegraphics[scale=0.3,keepaspectratio=true]{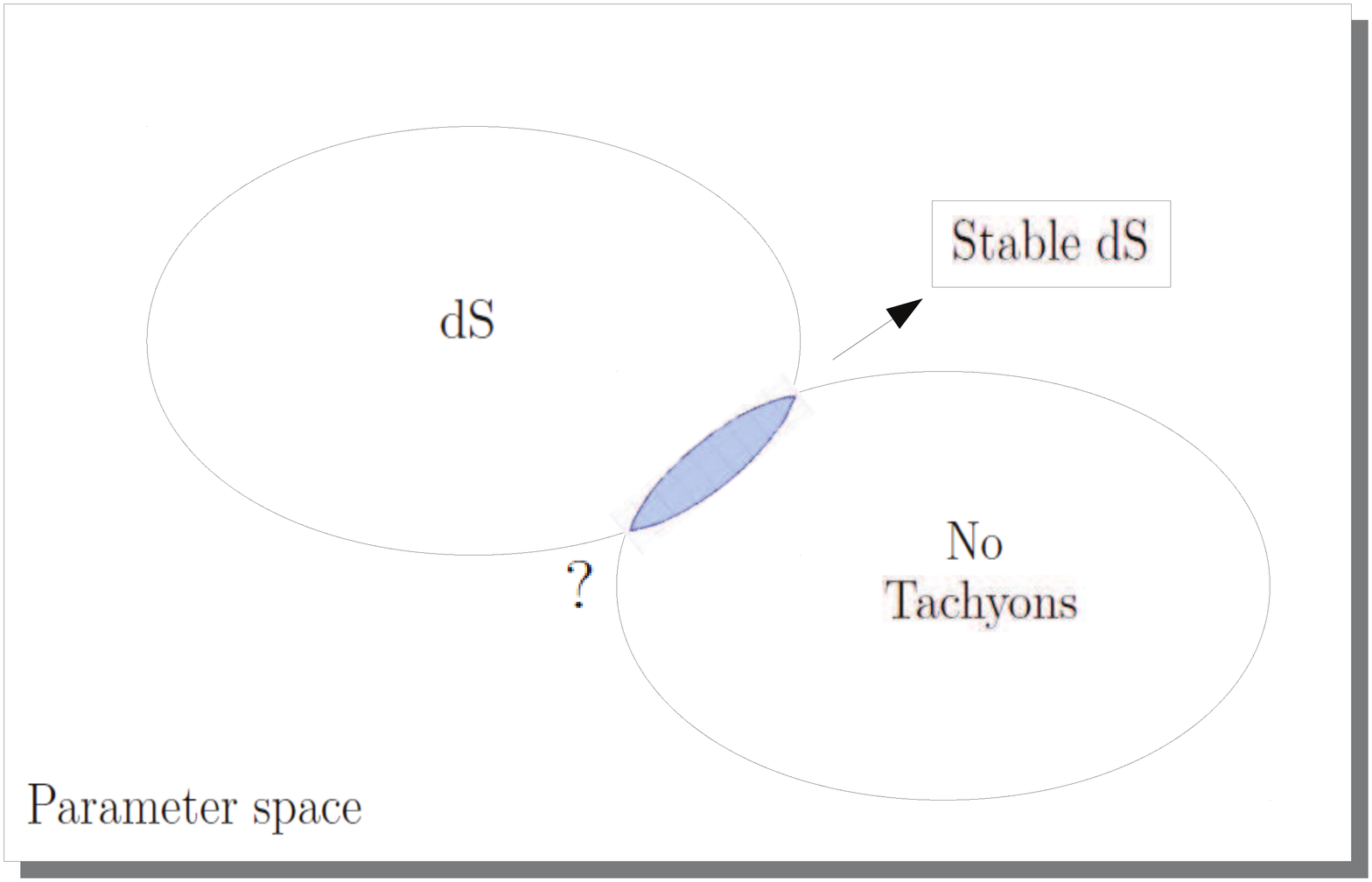} 
}
\caption{{\it The dS region as well as the tachyon-free region are subsets of positive measure in the $N$-dimensional parameter 
space of solutions. Generically, they will overlap in a region that may be small but still is expected to be of positive measure.}}
\label{fig:concept}
\end{center}
\end{figure}

A situation which then turns out to be of special interest is that of having exactly $2N$ fluxes turned on, so that one is able to write down the general solution to the field equations by expressing all the fluxes as a 
function of the supersymmetry breaking parameters.
The $N$-paramater family of solutions generated in this way should provide, at least as a matter of principle, enough freedom to have control upon the eigenvalues of mass matrix for the $N$ scalars retained
in the theory. Generically, this tachyon-free region in parameter space will have a non-trivial intersection with the dS region. This situation is conceptually depicted in figure~\ref{fig:concept}.

Our perspective points towards the possibility that, by introducing a non-isotropic setup admitting a $14$-dimensional space of solutions, one might have enough freedom for giving a positive mass to all the scalars in the $\textrm{SL}(2)^{7}$ model. 
At least qualitatively, increasing the number $N$ of fields should not spoil this picture.
The issue that we will address in the next section will be extimating how small the region of stable dS is compared to the whole parameter space of solutions. This will tell us how difficult it practically
becomes to scan the whole parameter space searching for stable dS vacua. This analysis, on the contrary, will certainly depend on $N$; in order to make predictions, we will now focus on the simpler $N=6$ case
corresponding with the isotropic model.

\subsection{Geometric IIA backgrounds: an over-constrained example}

Let us now consider the type IIA duality frame with O$6$-planes. This class of compactifications fits again 
the framework of $\cN=1$ effective descriptions in that there are two different kinds of allowed D$6$-branes, 
\emph{i.e.} those ones parallel to the O-planes which do not further break supersymmetry and those ones orthogonal to them 
which break supersymmetry down to $\cN=1$.  
In this context, in the isotropic limit, one can have the four different R-R gauge fluxes 
($\left\{F_{p}\right\}_{p\,=\,0,\,2,\,4,\,6}$), two independent components of $H$ flux, and three components of metric 
flux $\omega$, of which only two generate independent superpotential couplings (see footnote~\ref{c1_tilde}). 
The corresponding superpotential reads
\be
\label{W_geom_IIA}
W^{(\textrm{geom.~IIA})}\,=\,(a_0 - 3 \, a_1 \, U + 3 \, a_2 \, U^2 - a_3 \, U^3)\,+\,(b_0 - 3 \, b_1 \, U )\, S \,+\, 3 \, T \, (c_0 + c_{1} \, U)\ ,
\ee
where the type IIA interpretation of the above couplings can be found in table~\ref{table:unprimed_fluxes}. In this case, setting $\tilde{c}_{1}={c}_{1}$ as discussed in footnote~\ref{c1_tilde} has a special geometric interpretation, because it represents the correct way of solving the constraints $\omega^{2}=0$ and $\omega\,H=0$ which ensure the closure of the exterior derivative defined by the corresponding NS-NS background.

We choose the origin of moduli space to be
\be
\label{origin}
\begin{array}{cccccccc}
S_{0} & = & T_{0} & = & U_{0} & = & i & .
\end{array}
\ee
By applying the counting introduced in section~\ref{subsec:counting}, one immediately realises that the amount $M=8$ of fluxes is not large enough for following the above prescription for solving 
the equations of motion in the origin. After introducing the six supersymmetry breaking parameters by solving the \eqref{SUSY_break}, one finds 
\be
\label{lim_SUSY_IIA}
\begin{array}{lclc}
a_{2} & = & \frac{1}{9} \, \left(- B_{1} - B_{2} + 2B_{3}\right) & , \\[2mm]
a_{3} & = & \frac{1}{9} \, \left(- 3A_{1} - 3A_{2} + 2A_{3}\right) & , \\[2mm]
b_{0} & = & \frac{1}{9} \, \left(- 6A_{1} + 3A_{2} + A_{3}\right) & , \\[2mm]
b_{1} & = & \frac{1}{9} \, \left(2B_{1} - B_{2} - B_{3}\right) & , \\[2mm]
c_{0} & = & -\frac{1}{9} \, \left(3A_{1} + A_{3}\right) & , \\[2mm]
c_{1} & = & -\frac{1}{3} \, \left(B_{1} + B_{3}\right) & . \\[2mm]
\end{array}
\ee
After solving two axionic equations of motion by $a_{0}=a_{1}=0$, one is left with the following field equations
\be
\hspace{-4mm}
\label{iso_eom}
\begin{array}{rccc}
-3 A_{1}^2+2 A_{1} A_{3}-2 A_{2} A_{3}+A_{3}^2-3 B_{1}^2-2 B_{1} B_{3}-2 B_{2} B_{3}+B_{3}^2 & = & 0 & , \\[2mm]
9 A_{1}^2-6 A_{1}(A_{2}+A_{3})-2 A_{2} A_{3}+3 A_{3}^2+9 B_{1}^2-6 B_{1} B_{2}-6 B_{1} B_{3}-6 B_{2} B_{3}+3 B_{3}^2 & = & 0 & , \\[2mm]
9 A_{1}^2-9 A_{1}A_{2}+3 A_{1} A_{3}+3 A_{2} A_{3}+3 B_{1}^2-3 B_{1} B_{2}-B_{1} B_{3}-B_{2} B_{3}-4 B_{3}^2 & = & 0 & , \\[2mm]
6 A_{1} B_{1}-3 A_{1} B_{2}+3 A_{1} B_{3}-3 A_{2} B_{1}+3 A_{2} B_{3}+A_{3} B_{1}+A_{3} B_{2} & = & 0 & . 
\end{array}
\ee

\begin{figure}[h!]
\begin{center}
\begin{tabular}{cc}
\includegraphics[scale=0.8,keepaspectratio=true]{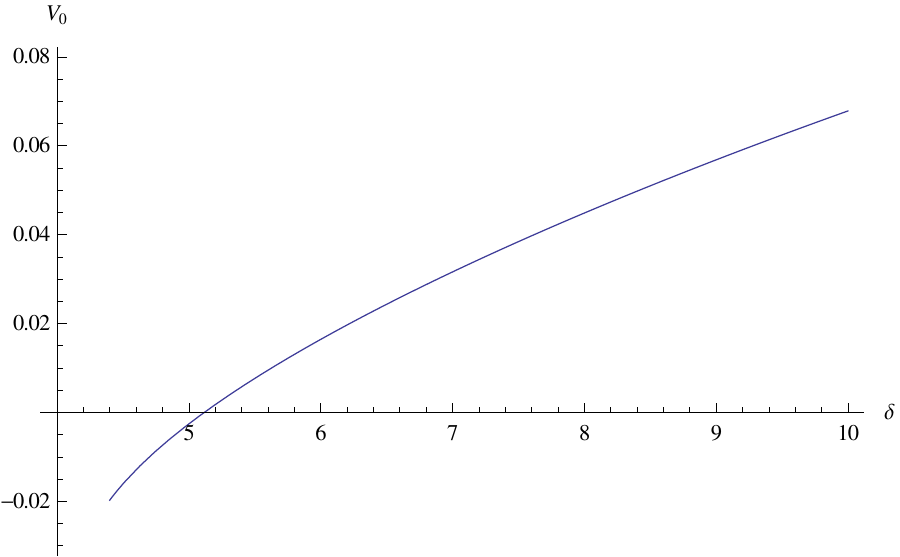}  &  \includegraphics[scale=0.8,keepaspectratio=true]{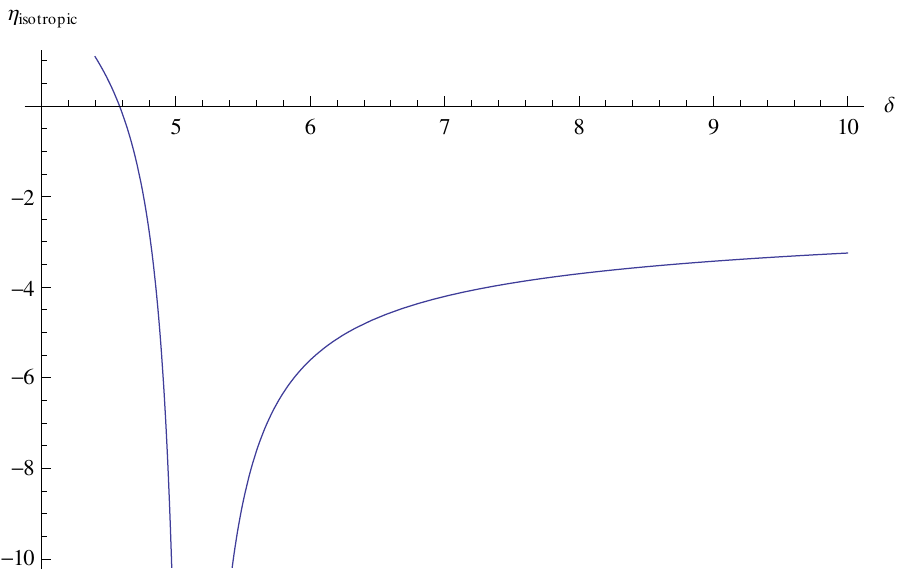}
\end{tabular}
\caption{{\it \emph{Left:} the one-parameter branch of dS solutions of geometric type IIA isotropic flux compactifications. The value of the energy $V_{0}$ is given as a function of the scan parameter $\d=\exp(A_{1})$.
The curve crosses the Minkowski line at $\d\sim 5.127$. \newline
\emph{Right:} the value of the $\eta$ parameter within the isotropic sector as a function of the scan parameter $\d$ already reveals the presence of tachyons
all along the dS branch.}}
\label{fig:dS_1}
\end{center}
\end{figure}

\begin{figure}[h!]
\begin{center}
\includegraphics[scale=0.8,keepaspectratio=true]{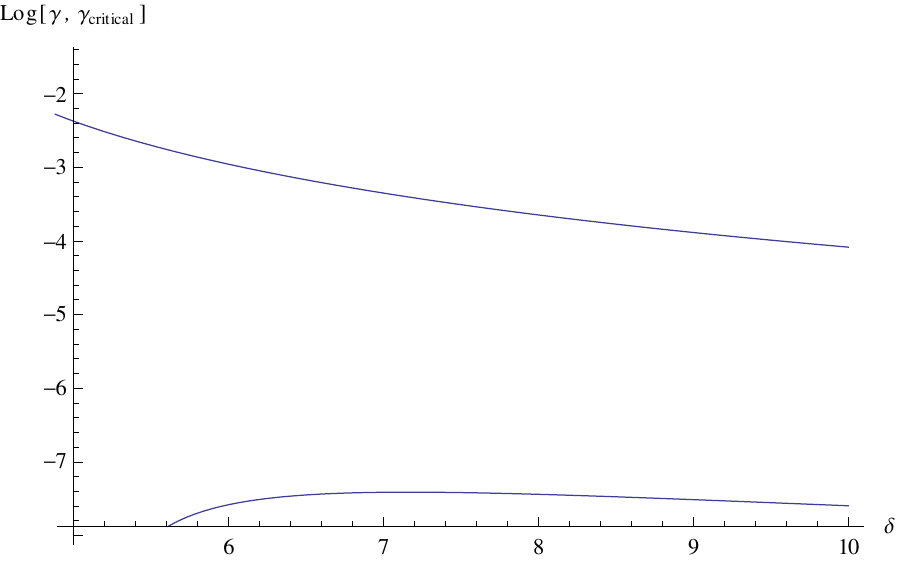} 
\caption{{\it Comparison between the value of $\g$ along the dS line and $\g_{\textrm{critical}}=\frac{n_{\textrm{eff}}}{3}-1$, \emph{i.e.} the maximal value of $\g$ for which the averaged sGoldstino mass
is not negative.}}
\label{fig:dS_2}
\end{center}
\end{figure}

Based on a technique from algebraic geometry called prime ideal decomposition and its implementation in the software \textsc{\,Singular\,} \cite{DGPS}, one can decompose the system \eqref{iso_eom} into its prime ideals which can be solved
separately and give rise to disconnected branches of the general solution. Only one of these branches happens to contain dS extrema and these organise themselves into a one-parameter family of solutions. 
This dS line is plotted in figure~\ref{fig:dS_1} together with the value of the $\eta$ parameter as a function of the scan parameter, which shows the presence of tachyons at every point. 
On the other hand, this over-constrained situation already suggests that one might not be able to tune the masses of the six isotropic real scalars and get stable dS points.
Further in figure~\ref{fig:dS_2} we show the normalised energy $\g$ and its corresponding critical value saturating the 
sGoldstino bound in \eqref{new_sG_bound}. This analysis manifests that the averaged sGoldstino (over real and imaginary parts)
is never tachyonic all along the dS, with $\g$ that stays orders of magnitude away 
from the critical value and only approaches it in the singular regime in which $\d$ goes to 
infinity\footnote{For further discussion on the presence of tachyons in these models, and the $\d$ limit in particular, see ref.~\cite{Danielsson:2012et}.}.

\section{Towards stable dS solutions}
\label{sec:non_geom_IIB}

In the previous section we have reviewed the case of geometric type IIA compactifications as an over-constrained example to which one can think of applying the method presented in section~\ref{subsec:counting}.
Now we would like to move to the most interesting situation in which $M$ exactly equals $2N$. In our isotropic setup ($N=6$) this implies the necessity of having $12$ fluxes. By taking a look at tables~\ref{table:unprimed_fluxes} and \ref{table:primed_fluxes},
one realises immediately that this requires including non-geometric fluxes both in type IIA and in type IIB. 

\subsection{A type IIB setup with generalised fluxes}
\label{subsec:non_geom_IIB} 

The setup we focus on is the IIB duality frame with O$3$ and O$7$-planes with generalised fluxes defining a locally geometric background, \emph{i.e.} $F_{3}$, $H$, $Q$ and $P$ fluxes. The induced superpotential is the one given in \eqref{W_IIB_non_geom}.
The aim of this section will be solving the equations of motion in the origin \eqref{origin} for the isotropic scalars in order to find the most general set of extrema in this case and study the stability features thereof. 

We will analyse here the special sub-case of a generalised type IIB background with only $F_{3}$, $H$ and $Q$ fluxes and vanishing $P$ flux. The general superpotential in this case is given in \eqref{W_FHQ}. This choice is motivated by the fact that, in this non-geometric setup, 
a dS critical point was found in ref.~\cite{deCarlos:2009fq} with the interesting property of being stable with respect to the six isotropic scalars. This supergravity solution, though, reveals the presence of tachyons in the non-isotropic directions of the scalar manifold. 
We will now solve the equations of motion for the scalars in this setup and show that the solutions span a parameter space of dimension six, in which the stable dS point of~\cite{deCarlos:2009fq} sits. 

In general, given such a $6$-dimensional parameter space, one can only hope to have control on the eigenvalues of the mass matrix in the six isotropic directions and there 
is no freedom left to stabilise the extra $8$ non-isotropic scalars which could then remain tachyonic.

The method for finding the general exact solution to the field equations in this non-geometric type IIB setup is presented in appendix~\ref{app:solutions}.

\subsection{Statistical scan and tests of RMT}
\label{subsec:RMT}

In the previous section we have motivated our interest in the study of non-geometric type IIB compactifications with $F_{3}$, $H$ and $Q$ fluxes. This seems to be the minimal setup for producing stable dS
solutions at least for what concerns the mass spectrum within the isotropic sector. At least qualitatively we have argued that this situation in which $N=6$ fields are retained can be regarded as a toy example
for extrapolating some universal features of the probability distribution of stable critical points for potentials with $N$ fields in general.

Analysing the probability distribution of stationary points of different types (\emph{i.e.} minima, maxima or saddle points) for classes of scalar potentials is equivalent to studying the distribution of the eigenvalues
of its Hessian matrix evaluated at those given points. In the case in which one assumes the scalar potential to be a random function, its Hessian matrix will be a symmetric matrix with random entries.

The distribution of the eigenvalues of a random matrix is the subject studied by Random Matrix Theory (RMT). This framework has been used in refs~\cite{Aazami:2005jf, Marsh:2011aa, Chen:2011ac, Bachlechner:2012at} in order to predict the heavy
suppression of stable dS vacua in string-inspired supergravity models with $N$ large enough.

Let us now go back to our general solutions presented in section~\ref{sec:non_geom_IIB} and appendix~\ref{app:solutions}. Even though this set of exact solutions is obtained analytically, one immediately realises
that scanning a parameter space of dimension six in order to find stable vacua requires a different approach, since the analytical beahviour of the eigenvalues of the mass matrix as a function of these parameters
is out of reach. 

\begin{figure}[h!]
\begin{center}
\includegraphics[scale=0.4,keepaspectratio=true]{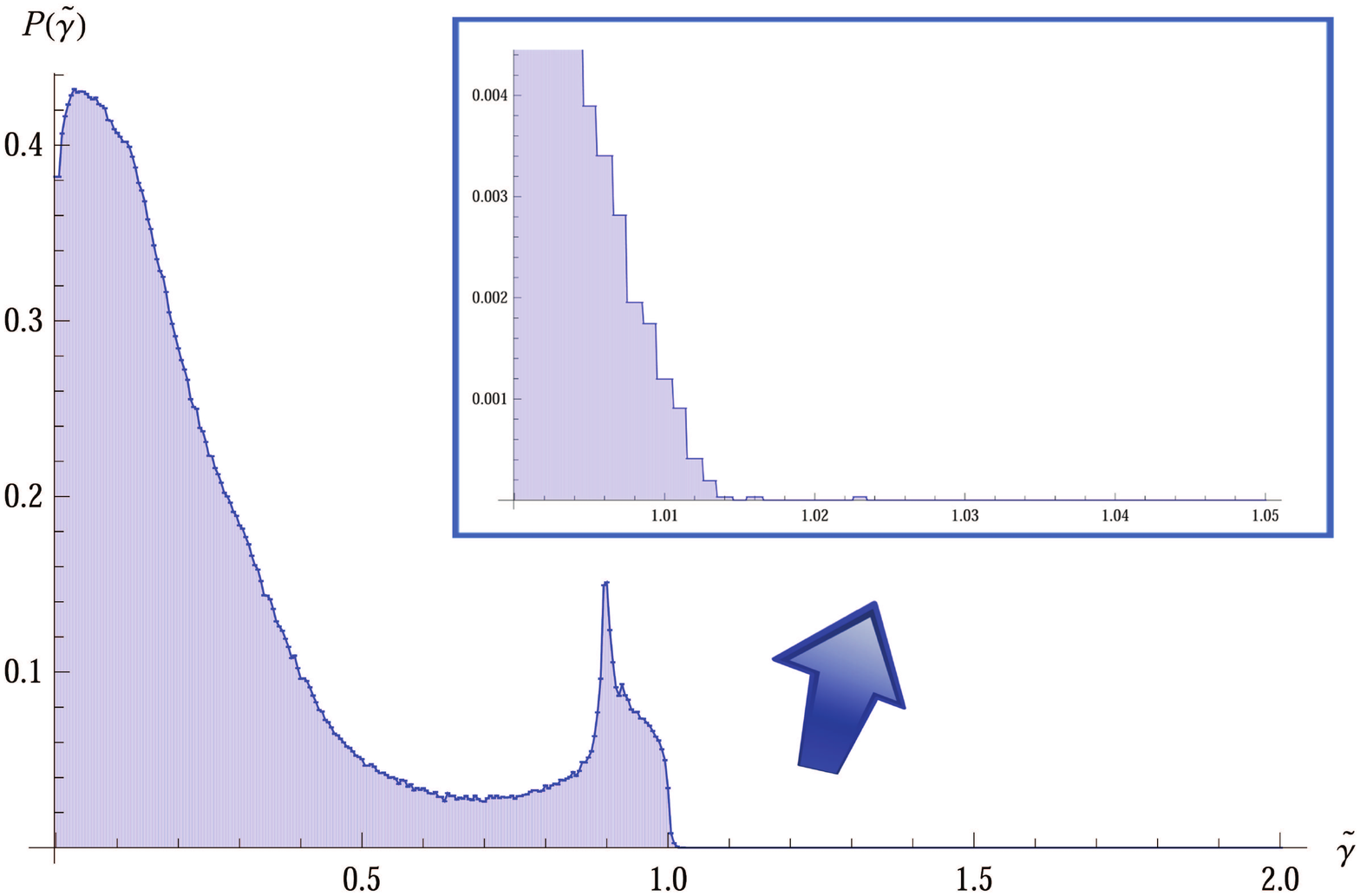} 
\caption{{\it Distribution of the fraction of tachyon-free critical points of the isotropic scalar potential with $N=6$ fields as a function of the uplift parameter 
$\tilde{\g}\,\equiv\,\frac{|DW|^{2}}{3|W|^{2}}$. Data produced from the analysis of $\,10^{7}$ random points. The zoomed diagram in the upper-right corner shows the presence of a stable dS tail with 
$1<\tilde{\g} \lesssim 1.02$. This tail was analysed using a sample bigger by a factor of $\,10$.}}
\label{fig:probability_distr_6}
\end{center}
\end{figure}

What we did was generating a statistically significant sample of random values for the six supersymmetry breaking parameters $\left\{A_{\a},\,B_{\a}\right\}̣_{\a\,=\,S,T,U}$, thus yielding an equally
big sample of random points in the space of solutions. In figure~\ref{fig:probability_distr_6} we show the resulting probability distribution of tachyon-free critical points as a function of the scanning
parameter\footnote{Please not that $\tilde{\g}=\g+1$, $\g$ being the normalised energy parameter introduced in the context of the sGoldstino analysis.} $\tilde{\g}$ representing the ratio between the positive contribution to the cosmological constant and the AdS scale. 

This result shows the presence of a peak of tachyon-free points in the AdS region with small $\tilde{\g}$ but away from the supersymmetric point at $\tilde{\g}=0$. This could be explained by observing
that the mass spectrum in the supersymmetric case is completely determined by the masses $m_{\a\b}$ of the chiral fermionic superpartners of the scalars. When these are massless, the corresponding partener
scalars sit at the conformal value $m^{2}=-\frac{2}{3}$. Non-zero fermionic masses will produce linear corrections to the off-diagonal elements of the scalar mass matrix and quadratic ones to its diagonal entries.
By studying in detail the sign of the eigenvalues, one finds the presence of at least one tachyon (though obviously still above the BF bound) as a generic feature, whereas introducing supersymmetry breaking
makes it slightly easier to arrange for all positive eigenvalues.

Furthermore, there appears a second sharp peak in figure~\ref{fig:probability_distr_6} in the AdS region but with values of $\tilde{\g}$ close to $1$. We do not have a clear explanation for this at the moment,
 though it seems to indicate that there might be a natural way of saturating the sGoldstino bound for the $\eta$ parameter in the AdS region for a certain value of $\g$.

Finally, we observe a tiny tail of stable dS points with the very specific feature of 
having $\tilde{\g}$ slightly above $1$. This seems to be in line with the results in refs~\cite{Sumitomo:2012wa, Sumitomo:2012vx, Sumitomo:2012cf}
where the relation between flux-induced potentials and RMT predictions is discussed, 
the final outcome being that string compactifications strongly suppress stable dS vacua with 
large values of the cosmological constant. 

In what follows, we would also like to make a comparison between our present analysis and 
the RMT-based results. It is important to keep in mind that our 
supergravity models describing a 
background with generalised fluxes only allows for a small number of 
random parameters that equals the number of diagonal entries of the mass matrix, 
whereas all the remaining off-diagonal ones are determined 
in terms of these by the equations of motion. In other words, 
while the RMT is dealing with a more or less arbitrary
random superpotential, we are considering a constrained string inspired subset.

We do not expect the RMT description to be applicable when the supersymmetry breaking scale 
and the AdS scale are of comparable size ($\tilde{\g}\gtrsim 1$). However, it is interesting 
to make a comparison in the far-right region where $\tilde{\g}$ is 
significantly larger than the Minkowski value of $1$. 
Using the estimate \cite{Borot:2010tr, Chen:2011ac}
\be
{\cal P} \,=\, \exp\left[-\frac{\log 3}{4}\,N^{2}+\frac{\log(2\sqrt{3}-3)}{2}\,N-\frac{\log N}{24}-0.0172\right] \ ,
\ee
one finds an estimated fraction $4.36\times 10^{-6}$ of stable dS for $N=6$. 
This would predict roughly $25$ to $30$ stable solutions out of the total 
of $\sim 6.3 \times 10^{6}$ dS solutions with $\tilde{\g}>1.5$ in our scan.
In contrast, we were unable to find a single example of stable 
dS in this region. This significant deviation
from the RMT prediction confirms our expectation that there is actually no regime in 
which our $\cN=1$ supergravity potentials, inspired by non-geometric string backgrounds, fit 
an RMT approximation. 

If one wants to reproduce the RMT predictions for stable dS vacua in the high $\tilde{\g}$ 
region, it is clear that one needs to include many more random parameters. The stringy 
origin of these remains, however, an open issue.

As a motivation towards further work in this direction, we would like to give a first look at the $N=14$ case corresponding to studying the full $\textrm{SL}(2)^{7}$ model relaxing isotropy. The analgous
setup allowing for enough freedom in parameter space will now have $28$ fluxes. By taking a look at the last column of table~\ref{table:unprimed_fluxes}, one observes that also in this non-isotropic situation
one needs to consider non-geometric backgrounds in any duality frame. As an example of this, we chose a type IIB background with the most general gauge fluxes ($F_{3}$ and $H$) plus those $Q$ flux components
appearing in superpotential terms with up to linear dependence on $U_{i}$ (The first two types of $Q$ flux in table~\ref{table:unprimed_fluxes}). These fluxes turn out to be exactly $28$ in total.
  
\begin{figure}[h!]
\begin{center}
\includegraphics[scale=1.1,keepaspectratio=true]{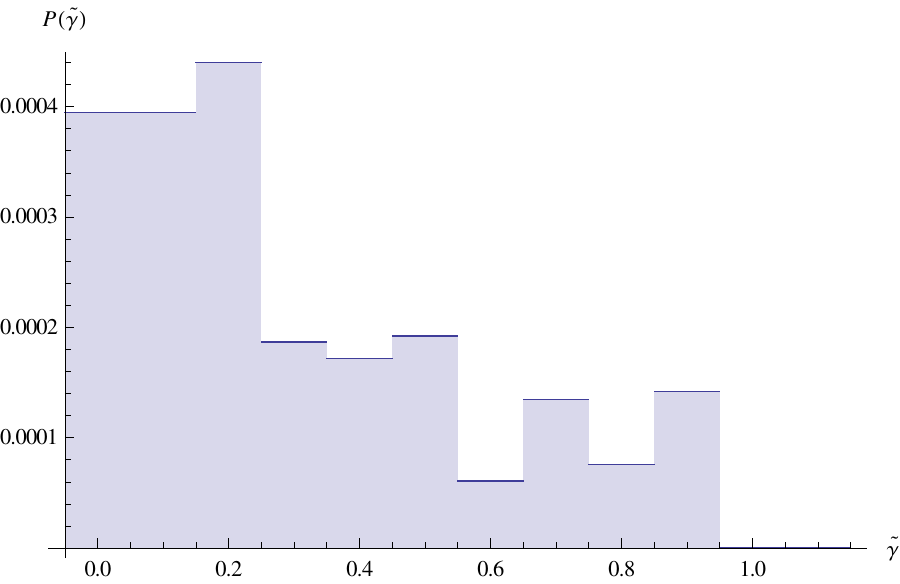} 
\caption{{\it Distribution of the fraction of tachyon-free critical points of the full scalar potential with $N=14$ fields as a function of the uplift parameter 
$\tilde{\g}\,\equiv\,\frac{|DW|^{2}}{3|W|^{2}}$. Data produced from the analysis of $\,5\times10^{5}$ random points. Although the statistical significance is not sufficient for detecting any stable 
Minkowski or dS tail, the general features in the AdS region are compatible with the results obtained for $N=6$.}}
\label{fig:probability_distr_14}
\end{center}
\end{figure}

In figure~\ref{fig:probability_distr_14} we give an idea of the results for the random scan in this case. These preliminary results were
obtained by means of a limited sample of data and we leave a detailed analysis for future purposes. Such a study will certainly be needed in order to investigate the presence of fully stable dS vacua, if any.
However, the present diagram of the probability distribution of tachyon-free points w.r.t. all real moduli roughly shows some common features with the $N=6$ case in the AdS region. 

\subsection{Analysis of the stable dS region}
\label{subsec:plots}

Now we would like to carry out a complementary analysis w.r.t. the statistical scan presented in the previous section. The aim is that of charting the region of stable dS inside parameter space in the 
IIB case previously introduced and studied. In this way, one can think of explicitely construct the concrete form of the conceptual picture given in figure~\ref{fig:concept} which applies in this case. 
 
\begin{figure}[h!]
\begin{center}
\scalebox{0.5}[0.5]{
\begin{tabular}{ccc}
\includegraphics[scale=1.1,keepaspectratio=true]{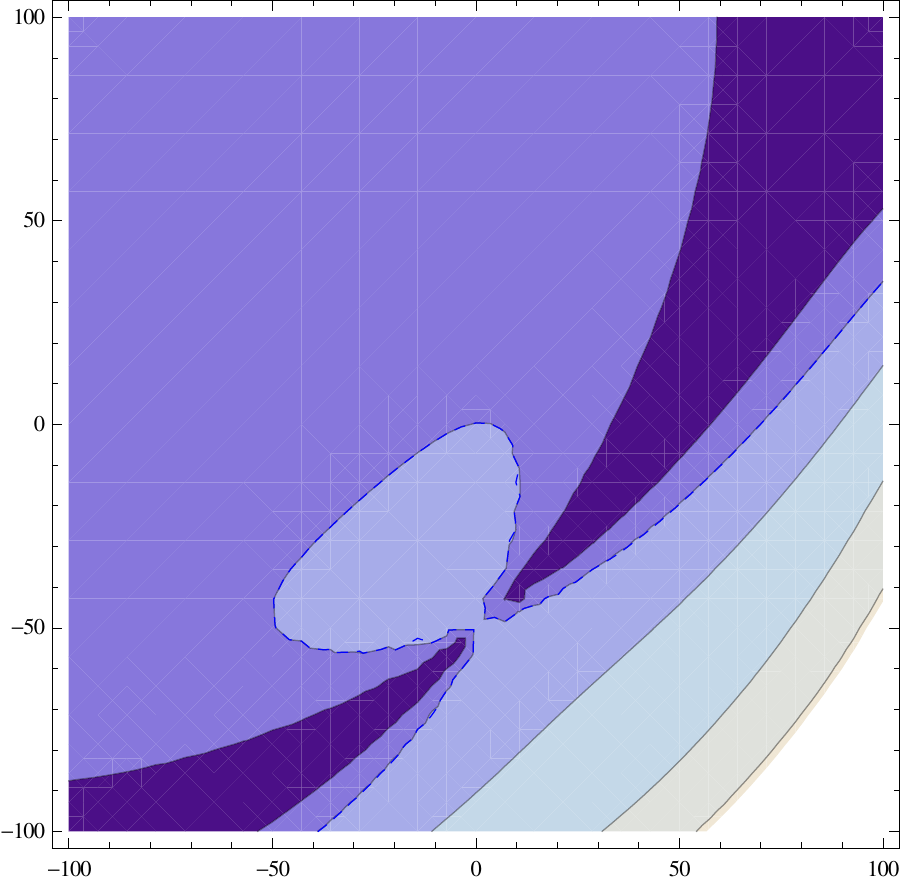} & \includegraphics[scale=1.1,keepaspectratio=true]{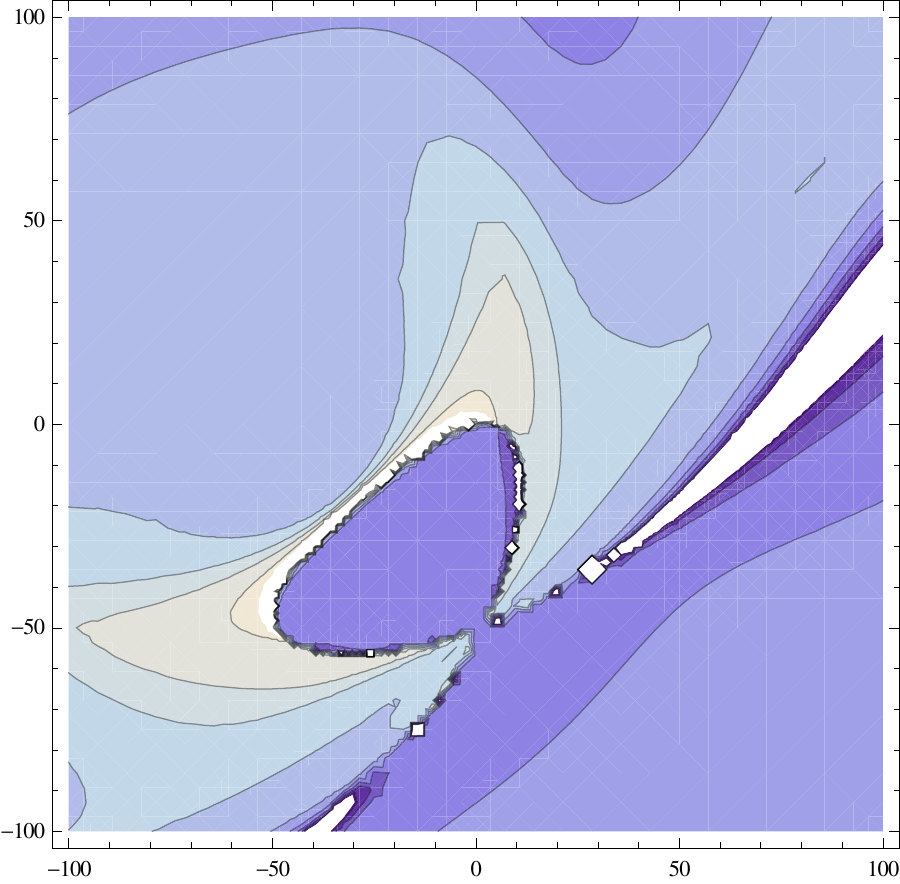} & \includegraphics[scale=1.1,keepaspectratio=true]{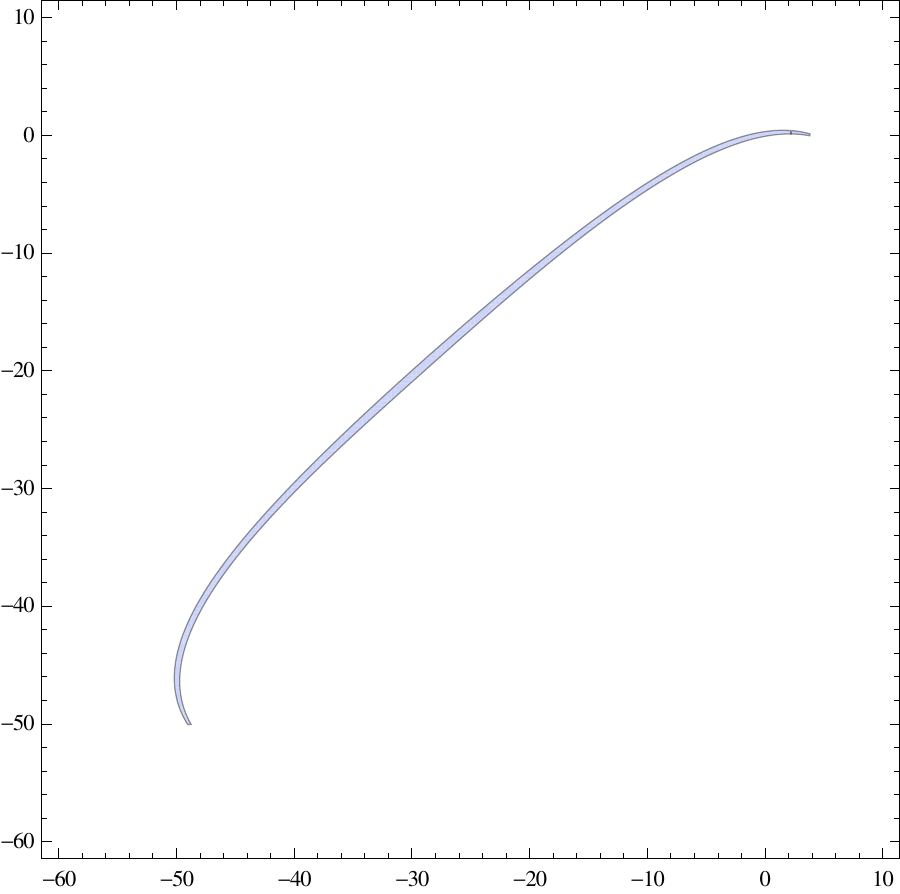}
\end{tabular}
}
\caption{{\it The parameter space of solutions projected on the $\left(A_{1},B_{1}\right)$ plane. \emph{Left:} Level curves of the cosmological constant; the lower-right corner filled with lighter colours and delimited
by the Minkowski (blue dashed) line corresponds to the dS region.  
\emph{Middle:} Level curves of the $\eta$ parameter; the tachyon-free region corresponds to the upper-left part filled with lighter colours.
\emph{Right:} The tiny region of overlap in parameter space corresponding with stable dS is zoomed in here.}}
\label{fig:stable_dS1}
\end{center}
\end{figure}

\begin{figure}[h!]
\begin{center}
\scalebox{0.5}[0.5]{
\begin{tabular}{ccc}
\includegraphics[scale=1.1,keepaspectratio=true]{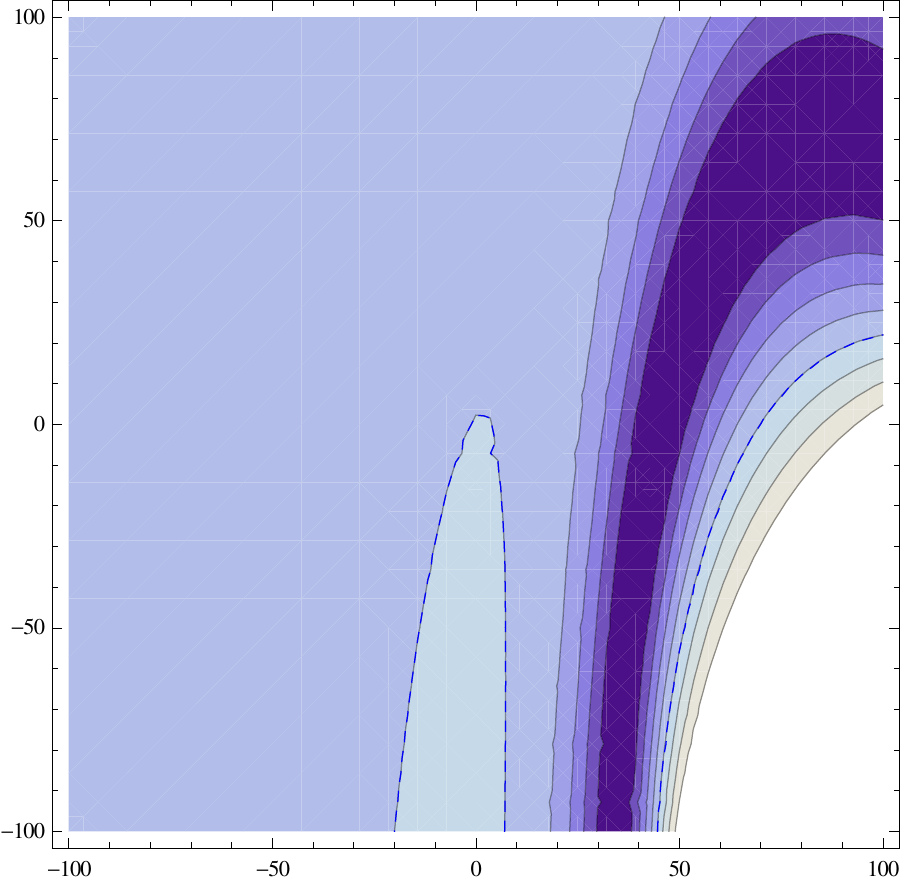} & \includegraphics[scale=1.1,keepaspectratio=true]{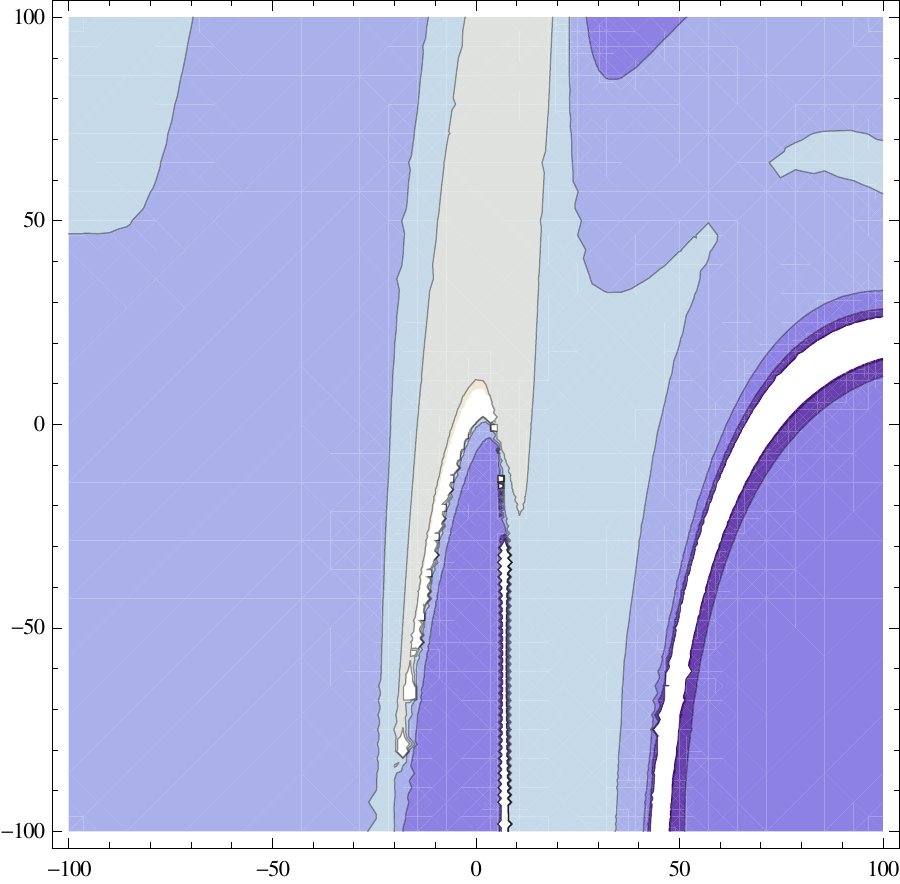} & \includegraphics[scale=1.1,keepaspectratio=true]{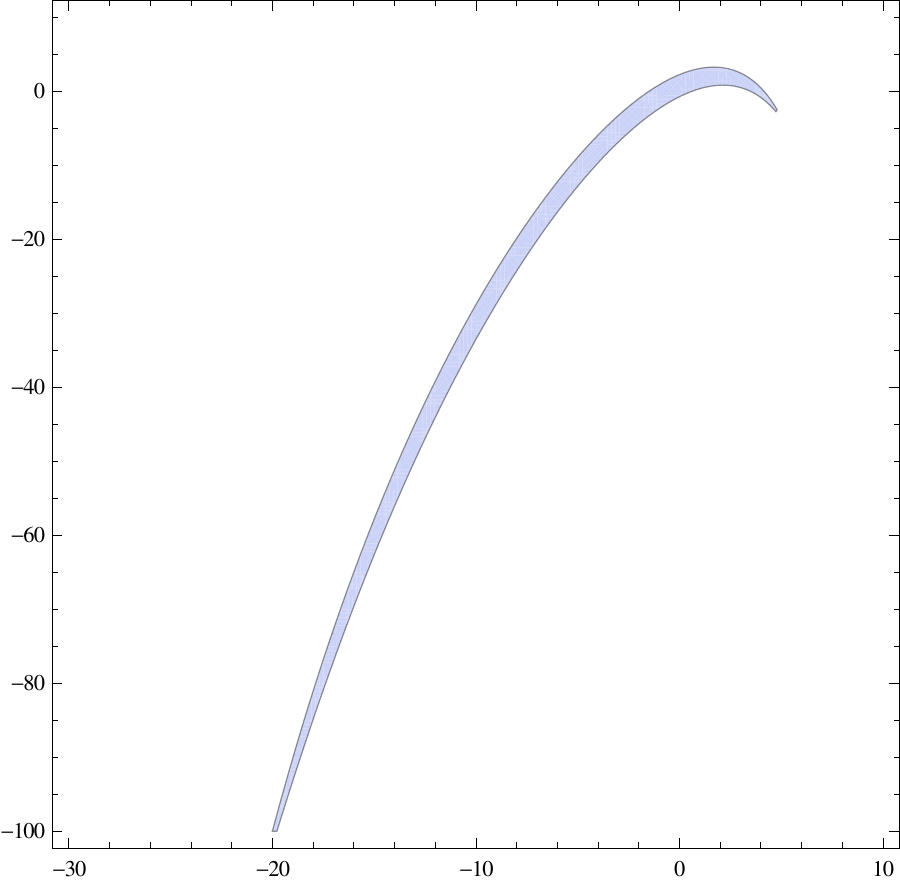}
\end{tabular}
}
\caption{{\it The parameter space of solutions projected on the $\left(A_{1},A_{2}\right)$ plane. \emph{Left:} Level curves of the cosmological constant; the dS regions are the ones filled with lighter colours below
by the Minkowski (blue dashed) lines.  
\emph{Middle:} Level curves of the $\eta$ parameter; the tachyon-free regions are filled with lighter colours.
\emph{Right:} The tiny region of overlap in parameter space corresponding with stable dS is zoomed in here.}}
\label{fig:stable_dS2}
\end{center}
\end{figure}

Since the parameter space of solutions has dimension six, we studied different two-dimensional slices thereof. We have fixed four out of six supersymmetry breaking parameters to the values corresponding to
the stable dS solution in ref.~\cite{deCarlos:2009fq} and moved about in the remaining two-dimensional slice. We have constructed the level curves of the scalar potential around such a point and the level
curves of the minimal eigenvalue of the mass matrix. Finally, we have studied the overlap between the dS region and the tachyon-free region thus locating stable dS points in parameter space.
The results of such an analysis are collected in figures~\ref{fig:stable_dS1} and \ref{fig:stable_dS2}, which represent the $\left(\Re(D_{S}W),\Im(D_{S}W)\right)$-plane and the $\left(\Re(D_{S}W),\Re(D_{T}W)\right)$-plane respectively.

The striking peculiarity of the dS region and the tachyon-free region is that they happen to be almost each other's negative and this also holds for all the other two-dimensional slices that are not given
here. This always makes the corresponding overlap region look like a very narrow band. By moving away from the stable dS point in which the plots are centered, the stable dS band tends to persist. This may
be regarded as an indication for the stable dS region to have the shape of a thick sheet in six dimensions, in which one of the dimensions is much narrower than the other ones. 
Studying the topology of such a stable dS region and in particular, as a first thing, understanding whether it consists of or more connected components, requires some further detailed investigation that we leave
for future work.

Furthermore, this complementary analysis corroborates what the random analysis was already suggesting, \emph{i.e.} that the very tiny stable dS regions always occurs very close a Minkowski critical point and hence
with values of $\tilde{\g}$ slightly above 1.

\section{Conclusions}
\label{sec:conclusions}

In this paper we have offered a framework for systematically analysing the
space of solutions of type II string
compactifications compatible with $\cN=1$ supersymmetry. The supergravity
effective description
realises string dualities as global symmetries, and therefore it comprises
different descriptions in different
duality frames in a single universal formulation.

Supersymmetry breaking parameters allow one to predict the dimensionality of
the space of solutions by means of a
very simple counting argument. In this context we were able to revisit the
case
of geometric type IIA compactifications. We find that the space of solutions
does not seem to contain enough degrees
of freedom for getting rid of the tachyons even within the isotropic sector.

Subsequently, we applied this novel prescription to investigate less
constrained situations
such as isotropic IIB compactifications with generalised fluxes. A
statistical scan as well as a charting of two-dimensional
slices of the parameter space reveal the presence of thin sheets of stable
dS regions with small cosmological constant
(compared to the AdS scale). Generically, since our parameter space is just
big enough for controlling
the eigenvalues of the mass matrix in the isotropic sector, we expect that
the remaining non-isotropic sector will contain
tachyons. This we have checked in some explicit cases, even though we have
not performed a full analysis.

However, by considering non-isotropic setups with a larger parameter space,
our general expectation is that there will
exist a non-empty stable dS region. This region will probably be
considerably thinner
than its isotropic counterpart. Indeed, a first random scan of the
non-isotropic case (although not with enough
statistical significance), has exhibited no examples of tachyon-free dS
critical points.
This means that the issue of the existence of a fully stable dS vacuum,
w.r.t. to all fourteen non-isotropic moduli, still
remains to be addressed. We hope to come back to this point in the future.

Our work shows that stable de Sitter vacua are much less frequent than one
naively would expect. In particular, we
have observed the necessity to add at least some flux that is non-geometric
from both the type IIA and the type IIB point of view.
Furthermore, our analysis hints at tantalizing conspiracies, which suggest
that dS vacua, if at all present, have unnaturally small
values (compared to the AdS scale). It would be interesting to analytically
understand the reason for these conspiracies, and
to investigate their possible phenomenological consequences.

%
%

\section*{Acknowledgments}

We would like to thank Johan Bl\r{a}b\"ack, Adolfo Guarino, Joseph Minahan, Oscar Varela for stimulating discussions. We are also especially greatful to Gary Shiu, Thomas Van Riet and Timm Wrase for valuable comments
on a draft version of this manuscript. The work of the authors is supported by the Swedish Research Council (VR), and the G\"oran Gustafsson Foundation.

\newpage

%
%

\appendix

\section{All the solutions of type IIB with $F$, $H$ and $Q$}
\label{app:solutions}

The general form of the superpotential is obtained by taking \eqref{W_all_fluxes} and switching of all the couplings but those ones in $P_{F}$, $P_{H}$ and $P_{Q}$. This yields
\be
\label{W_FHQ}
\begin{array}{lclcc}
W^{(P=0)} & = & a_0 - 3 \, a_1 \, U + 3 \, a_2 \, U^2 - a_3 \, U^3 + ( b_0 - 3 \, b_1 \, U + 3 \, b_2 \, U^2 - b_3 \, U^3 ) \, S  & + & \\[2mm]
& + & 3 \, T \, (c_0 + c_{1} \, U - c_{2} \, U^2 - c_3 \, U^3) &  & . 
\end{array}
\ee
After solving the ``broken'' supersymmetry conditions \eqref{SUSY_break} in terms of $\left\{b_{2},\,b_{3},\,c_{0},\,c_{1},\,c_{2},\,c_{3}\right\}$ via
\be
\label{lim_SUSY_IIB}
\begin{array}{lclc}
b_{2} & = & \frac{1}{6} \, \left(3a_{1} - a_{3} + 2b_{0} + A_{1} - A_{2}\right) & , \\[2mm]
b_{3} & = & \frac{1}{2} \, \left(-a_{0} + 3a_{2} + 6b_{1} - B_{1} + B_{2}\right) & , \\[2mm]
c_{0} & = & \frac{1}{12} \, \left(9a_{1} + a_{3} + 4b_{0} - A_{1} - A_{2} - 2A_{3}\right) & , \\[2mm]
c_{1} & = & \frac{1}{12} \, \left(9a_{0} - 15a_{2} - 12b_{1} - 3B_{1} - 3B_{2} - 2B_{3}\right) & , \\[2mm]
c_{2} & = & \frac{1}{12} \, \left(9a_{1} - 7a_{3} - 4b_{0} - 5A_{1} - A_{2} + 2A_{3}\right) & , \\[2mm]
c_{3} & = & \frac{1}{12} \, \left(-3a_{0} - 3a_{2} + 12b_{1} - 3B_{1} + B_{2} + 2B_{3}\right) & , \\[2mm]
\end{array}
\ee
one finds
\be
\begin{array}{l}
(6  B_{1}+3 B_{2}-3 B_{3}) a_{0} + 3 (6 A_{1}+3 A_{2}-A_{3}) a_{1} -9 (2 B_{1}+ B_{2}-B_{3}) a_{2} + \\
 - (6 A_{1}+3 A_{2}- A_{3}) a_{3}-8 A_{3}b_{0} +24 B_{3}b_{1} -6 A_{1}^2-A_{1} (3A_{2}+A_{3})-6 B_{1}^2 + \\
-3 B_{1} B_{2}-3 B_{1} B_{3}-A_{2}^2+A_{2} A_{3}-B_{2}^2+3 B_{2} B_{3}+2 A_{3}^2+2 B_{3}^2 \,=\, 0 \ , \\[2mm]
3 (2 A_{1}-A_{2}+A_{3}) a_{0} -3 (6 B_{1}-3 B_{2}+ B_{3}) a_{1} -9 (2A_{1} - A_{2}+ A_{3}) a_{2} -8 B_{3} b_{0} -24 A_{3} b_{1} + \\
+(6 B_{1} -3  B_{2}+ B_{3}) a_{3} - A_{1} (3 B_{2}+B_{3}) +3 B_{1} A_{2}+3 B_{1} A_{3}+ A_{2} B_{3}-3 B_{2} A_{3}\,=\, 0 \ , \\[2mm]
3 (3 B_{1}+4 B_{2}+3 B_{3}) a_{0} +9 (3 A_{1}+4 A_{2}+A_{3}) a_{1} -3 (9 B_{1}+12 B_{2}+B_{3})a_{2}  + \\
-(9 A_{1} +12 A_{2} -5 A_{3}) a_{3} + 8A_{3} b_{0} -24 B_{3}b_{1} +9 A_{1}^2 + A_{1} (-15 A_{2}+A_{3})+9 B_{1}^2  + \\
-15 B_{1} B_{2}+3 B_{1} B_{3}-4 A_{2}^2-A_{2} A_{3}-4 B_{2}^2-3 B_{2} B_{3}+2 A_{3}^2+2 B_{3}^2\,=\, 0 \ ,\\[2mm]
-9 (A_{1} +A_{3}) a_{0} + 9 (3 B_{1} + B_{3}) a_{1} + 3(9A_{1}+A_{3}) a_{2} -(9 B_{1}-5B_{3}) a_{3}  + \\
 +8 B_{3} b_{0} +24A_{3} b_{1} + A_{1} (3 B_{2}+B_{3}) -3 B_{1} A_{2}-3 B_{1} A_{3}-A_{2} B_{3}+3 B_{2} A_{3} \,=\, 0 \ ,\\[2mm]
-9(B_{1} -B_{2}-2B_{3}) a_{0} + 3(-3 A_{1}+3 A_{2}+2 A_{3}) a_{1} + 3(9 B_{1}- B_{2}-2 B_{3}) a_{2} + \\
+ (3 A_{1}+5 A_{2}-18 A_{3}) a_{3} + 8 A_{2}b_{0} -24B_{2} b_{1} -3 A_{1}^2 + A_{1} (-24 b_{0}+4 A_{2})-9 B_{1}^2  + \\
+ 72 B_{1} b_{1}+12 B_{1} B_{2}-A_{2}^2-4 A_{2} A_{3}-3 B_{2}^2-4 B_{2} B_{3}+4 A_{3}^2-4 B_{3}^2 \,=\, 0 \ ,\\[2mm]
3 (3 A_{1}-3 A_{2}-2 A_{3}) a_{0} -3(3B_{1}-3 B_{2}+10 B_{3}) a_{1} -3(9A_{1}- A_{2} +10 A_{3}) a_{2} + \\
+(3 B_{1}+5 B_{2}-6 B_{3}) a_{3} +8 B_{2} b_{0}+24 A_{2} b_{1}-A_{1} (-6 B_{1} +72 b_{1}+8 B_{2}) + \\
-24 B_{1} b_{0}+2 A_{2} B_{2}+4 A_{2} B_{3}-4 B_{2} A_{3}+8 A_{3} B_{3} \,=\, 0 \ .
\end{array}\nn
\ee
The above set of equations can be then solved with respect to $\left\{a_{0},\,a_{1},\,a_{2},\,a_{3},\,b_{0},\,b_{1}\right\}$ which only appear linearly. This procedure combined with \eqref{lim_SUSY_IIB} gives rise to
to the general solution expressing the $12$ fluxes as a function of the six supersymmetry breaking parameters. 

%
%

\small

\bibliography{references}
\bibliographystyle{utphys}

\end{document}